\documentstyle[12pt]{article}

\input{tcilatex}
\begin{document}

\noindent {\Large {\bf Ion beam sputtering induced ripple formation in thin
metal films }} \vspace{0.7in}\newline

\begin{center}
{\bf P. Karmakar\footnote{{\bf Corresponding author. e-mail:
prasanta@surf.saha.ernet.in}} and D. Ghose } \vspace{0.1in} \\[0pt]
{\it Saha Institute of Nuclear Physics,\\[0pt]
Sector - I, Block - AF, Bidhan Nagar, Kolkata 700064, India} \vspace{0.4in}%
\\[0pt]
{\bf Abstract}\\[0pt]
\end{center}

We have observed the formation of ripples in a number of thin metal films,
e.g. $Au$, $Pt$, $Ag$, $Cu$ and $Co$ under $Ar^{+}$ ion beam sputtering at
grazing incidence. The structures are found to be quite stable under ambient
conditions. The results show that the ripple formation in polycrystalline
metallic films relies on the erosion-induced surface instability similar to
that in amorphous materials.


\pagebreak

\section{\noindent Introduction}

{One of the experimental techniques for fabricating submicron-sized
well-defined pattern} on solid surfaces is the low energy ion beam
sputtering. Among the various ion beam induced morphologies, the formation
of periodic ripple structures has attracted much research interest in recent
years both for understanding the underlying physical mechanisms [1] as well
as for potential applications in the field of nanotechnology, e.g. ripple
pattern might be exploited as a template for quantum dot formation [2].
Although a large number of works on the ripple morphology have been carried
out in semiconductor materials, little work has been done on metal surfaces
[1]. The group of Valbusa [3] was pioneer in observing nanoscale-ripples on
single crystals of $Cu$ and $Ag$, where they utilized the presence of
Erlich-Schwoebel (ES) barrier at crystalline step edges to develope ripples
along energetically favoured crystallographic directions. Since the
diffusion-biased ripples are highly sensitive to the substrate temperature,
the ripple morphology are found to be unstable at room temperature, because
of the low ES barrier heights [4]. Very recently, Sekiba et al. [5] have
shown that {\it in-situ} oxidation of the rippled surface immediately after
the formation may provide a long term stability at room temperature or
higher.

Polycrystalline metal thin films have wide industrial applications in
electronic, magnetic, and optical devices [6]. It will, therefore, be
interesting to explore the possibility of formation of correlated surface
features such as wavelength selected ripple structures, at sub-micometre
length scales, in these systems also. In this paper, we report the
development of ripple topography at grazing ion beam sputtering on a number
of thin metallic films at ambient temperature.

\section{Experimental}

Thin films of $Au$, $Pt$, $Ag$, $Cu$ and $Co$ were deposited by d. c.
magnetron sputtering (Pfeiffer, PLS 500) onto commercially available
polished $Si(100)$ wafers, previously degreased and cleaned. The base
pressure in the deposition chamber was $2\times 10^{-6}$ mbar. The film
thicknesses were in the range of $30$ to $200$ nm. These samples were then
sputtered with mass analyzed $5-27$ keV $Ar^{+}$ ions in a low-energy ion
beam (LEIB) system developed in the laboratory [7]. The angle of ion
incidence with respect to the surface normal was varied between $10^{0}$ and 
$80^{0}$. The beam current through a $4$ mm diameter aperture was in the
range of $2-3$ ${\mu }A$. The samples were exposed to ion doses in the range 
$5\times 10^{15}-$ $5\times 10^{17}$ $ions/cm^{2}$, which were measured by a
current integrator (Danfysik, model 554) after suppression of the secondary
electron emission. The base pressure in the target chamber was less than $%
5\times 10^{-8}$ mbar. The surface morphology of the ion-irradiated samples
was examined by a Park Scientific AFM (Auto Probe CP) in the contact mode.
All the measurements were carried out in air at room temperature.

\section{Results and discussion}

Atomic force microscopy of the as-deposited films shows that the initial
surface topography contains characteristic bump-like structures, which are
typical for thin films grown by the process of sputtering [8]. The evolution
of surface morphology as a function of the angle of ion incidence for a
given ion dose and energy shows first the development of mound structure
which tends to grow upto the angle of incidence $\theta $ $\sim $ $50^{0}$.
Further increase of the incidence angle shows the beginning of distinct
changes of the surface morphology which ultimately ends up with regular
ripple structures at grazing ion incidence. Figs. 1(a) - (t) show some
selected AFM images of the sputtered $Co$, $Cu$, $Ag$, $Pt$ and $Au$ films,
respectively, at the angles of $60^{0}$, $70^{0}$, and $80^{0}$, because at
these angles the gradual morphological transitions are clearly visible and
this is a general feature for all the metallic films studied in the present
sputtering conditions. At $60^{0}$ weakly pronounced ripples with wave
vector parallel to the ion beam, especially in $Co$ and $Cu$ substrates,
appear. At $70^{0}$ the morphology shows the development of arrays of tiny
cones aligned along the projection of the ion beam direction. Finally, at $%
80^{0}$ regular ripple-like surface instability with the wave vector
perpendicular to the ion beam direction is developed. In passing we mention
that the AFM images of the ripples including that of the other morphological
structures are found to be quite reproducible even after several weeks of
bombardment.

For the quantitative analysis of the ripple morphology, we have calculated
numerically the height-height autocorrelation function $C({\bf r})=$ $%
\langle [h({\bf r})h(0)]\rangle $, where $h({\bf r})$ is the relative
surface height at the position ${\bf r}$ and $\langle $ $\rangle $ denotes
an average over all positions and directions. As an illustration, Fig. 2
shows a typical AFM image of the rippled structure on a $Au$ film together
with the corresponding two-dimensional autocorrelation function. The ripple
wavelength $\Lambda $ is defined as the separation between the central peak
and the first secondary correlation maximum while taking linear scans of $C$%
. Fig. 3 shows a typical set of data for the ripple wavelength $\Lambda $ as
a function of the substrate material when sputtered with the same total ion
dose and energy.

Valbusa et al. [3, 4] showed that sputtering of metal surfaces involves two
types of surface instabilities depending on the angle of ion incidence, $%
\theta $, the first one arising from erosion process and the other deriving
from anisotropic surface diffusion. The erosion-induced surface instability
dominating at grazing incidence $\theta >\theta _{c}$ ($\theta _{c}\approx
50^{0}-70^{0}$ [9]) leads to ripple structures aligned parallel to the ion
beam projection, independent on the surface crystallinity or orientation.
The erosive regime is supposed to govern by the Bradley and Harper (BH)
theory [10], where{\ the surface height evolution }${h(}x,y,t{)}${\ can be
described as [10, 4]}

\begin{equation}
\frac{\partial h}{\partial t}=-v_{0}+\gamma \frac{\partial h}{\partial x}%
+\nu _{x}\frac{\partial ^{2}h}{\partial x^{2}}+\nu _{y}\frac{\partial ^{2}h}{%
\partial y^{2}}-K\nabla ^{4}h+\eta ,
\end{equation}
where $v_{0}$ is the surface erosion rate of the flat surface at normal
incidence, $\gamma $ is related to the derivative of the sputtering yield
with respect to the angle of ion incidence, $\nu _{x}$ and $\nu _{y}$ are
the effective surface tensions generated by the surface erosion process, the
constant $K$ is related to the surface diffusion which is activated by
different physical processes, namely, thermal or ion beam induced or both
[1] and, finally $\eta $ is the noise term associated with the randomness of
the bombarding ions.

Eq. (1) can further be extended to crystalline materials by including the
effects of anisotropic diffusion in different crystallographic directions as
well as the existence of the ES barrier at the step edges [11]. One of the
consequences of the presence of the ES barrier is that the ripple structure
in metal surfaces could be formed even at normal ion beam incidence ($\theta 
$ $\approx $ $0^{0}$) by tuning the surface temperature [11]. However, for
polycrystalline thin films, the grains are mainly randomly oriented and the
grain sizes are usually much smaller compared to the film thickness [12,
13]. For such a system, the existence of ES barrier is improbable because of
the lack of well-defined atomic steps at the surface [12]. Therefore, the
approximation of isotropic diffusivity as in amorphous materials seems to
hold good also in polycrystalline metallic films. The stability of the
ripple structure at room temperature indicates that the thermally activated
diffusion energy barriers in thin polycrystalline films is comparatively
higher than that in monocrystalline metal surfaces. Rossnagel and Robinson
[14] measured the activation energy for adatom surface diffusion on various
polycrystalline materials from the Arrhenius plot of the sputter cone
spacings for several temperatures. The barrier heights lie typically in the
range of $0.3-1$ eV. Similar plots for ripple wavelengths on $Ag$ single
crystals yields activation energy around $0.15$ eV [4].

At room temperature the surface diffusion is driven by the collisional
effects rather than pure thermal effect [15]. The mobility of adatoms is
believed to originate from the overlapping collision cascades due to
multiple ion impact [16]. Carter and Vishnyakov [17] proposed to add a
ballistic smoothening term of the form $\mid A(E,\theta )\mid \nabla ^{2}h$
in Eq. (1) in order to account the effect of recoiling-adatom diffusion
induced by ion irradiation at a given energy $E$. More recently, Makeev et
al. [1] showed that fourth-order derivatives of the surface height function $%
h(x,y)$ may also cause the smoothing effect which, however, does not involve
real mass transport. For such a case the wavelength of the ripples, in
present experimental conditions, can be derived as $\Lambda =2\pi \sqrt{%
\frac{2D_{yy}}{\mid \nu _{y}\mid }}$, where $D_{yy}$ is the ion induced
smoothing coefficient in the $y$ direction as defined in Eq. 52 of Ref. 1.
The ratio $D_{yy}/\nu _{y}$ can be estimated from the values of the ion
penetration depth $a$ and the longitudinal and lateral stragglings $\sigma $
and $\mu $ , respectively, using the computer code SRIM [18]. Although the
experimental data at fixed bombarding ion energy and dose follow the same
trend as the theoretically calculated $\Lambda $ in different substrate
elements, the theory underestimes substantially the experimental wavelength
values (cf., Fig. 3). The discrepancy is due to the fact that the ripple
wavelength $\Lambda $ is found to increases with the ion dose $\phi $ as a
power law $\Lambda \sim \phi ^{n}$ with the exponent $n=0.53$, e.g. measured
for $Pt$ films [19], whereas the BH model predicts no dependence of the
ripple wavelength on ion dose or sputtering time.

In passing we should mention that the ripple formation does not depend on
the initial surface topography, i.e. whether the initial surface is
atomically flat or rough, the surface is always characterized by ripples at
grazing ion beam sputtering. Such a result is due to the effect of the noise
term in Eq. (1), where random arrival of the bombarding ions destroys the
old surface morphology and generates new morphology which grows with the
erosion time [20].

Finally, we have also bombarded clean $Si(100)$ wafers in the angular range $%
10^{0}$ and $80^{0}$ under identical conditions. Here ripples are formed
only at the ion incidence angle of $60^{0}$ and the wave vector of the
ripples is found to be parallel to the ion beam direction (Fig. 4). The most
interesting observation, however, is that the $Si$ ripples are generated at
doses $>$ $10^{17}$ $ions/cm^{2}$ in agreement with that of others, e.g.
[21], in contrast to the metallic ripples which begin to develop at much
lower doses of about $10^{15}$ $ions/cm^{2}$.

\section{Conclusion}

In conclusion, grazing ion beam sputtering at room temperature can induce
ripple structures on thin metallic films similar to that reported on single
crystal metal surfaces at low temperature [4]. Unlike the latter, such
ripples are found to be quite stable at ambient conditions. The present
experiment also indicates that the ripples do not form, and the surface
undergoes kinetic roughening so long as $\theta <\theta _{c}$, where $\theta
_{c}\simeq 50^{0}-60^{0}$. Finally, it is noted that, independent of the
initial morphology of the surface, the ripples in metallic films start to
generate at ion doses as low as $10^{15}$ $ions/cm^{2}$ which is roughly two
orders of magnitude smaller than that for the ripples formed on $Si$
surfaces.

\section{Acknowledgement}

\ The authors thank Mr. A. Das for technical assistance during the AFM
measurements.

\pagebreak

\section*{References}

\begin{itemize}
\item[{[1]}]  M. A. Makeev, R. Cuerno, and A. -L. Barab\'{a}si, Nucl.
Instrum. and Meth. B 197 (2002) 185.

\item[{\lbrack 2]}]  G. Li, J. Zhang, L. Yang, Y. Zhang, and L. Zhang,
Scripta mater. 44 (2001){\bf \ }1945.

\item[{\lbrack 3]}]  U. Valbusa, C. Boragno, and F. Buatier de Mongeot,
Materials Science and Engineering C 23 (2003) 201.

\item[{\lbrack 4]}]  U. Valbusa, C. Boragno, and F. Buatier de Mongeot, J.
Phys.: Condens. Matter 14 (2002) 8153.

\item[{\lbrack 5]}]  D. Sekiba, S. Bertero, R. Buzio, F. Buatier de Mongeot,
C. Boragno, and U. Valbusa, Appl. Phys. Lett. 81 (2002) 2632.

\item[{\lbrack 6]}]  {\it International Technology Roadmap for Semiconductors%
} (Semiconductor Industry Association, San Jose, CA, 2001).
[http://public.itrs.net/Files/2001ITRS/Home.htm].

\item[{\lbrack 7]}]  P. Karmakar and D. Ghose, in: {\it Proc. DAE-BRNS
Indian Particle Accelerator Conference-2003}, Eds. S. C. Bapna, S. C. Joshi
and P. R. Hannurkar (Allied Publishers Pvt. Ltd., New Delhi, 2003), p. 169.

\item[{\lbrack 8]}]  G. S. Bales, R. Bruinsma, E. A. Eklund, R. P. U.
Karunasiri, J. Rudnick, and A. Zangwill, Science {\bf 249}, 264 (1990).

\item[{\lbrack 9]}]  R. M. Bradley and J. M. E. Harper, J. Vac. Sci.
Technol. A 6 (1988) 2390.

\item[{\lbrack 8]}]  P. Sigmund, J. Mater. Sci. 8 (1973) 1545.

\item[{\lbrack 9]}]  G. Costantini, Thesis, University of Genova, December
1999.

\item[{\lbrack 10]}]  R. M. Bradley and J. M. E. Harper, J. Vac. Sci.
Technol. A 6 (1988) 2390.

\item[{\lbrack 11]}]  G. Costantini, S. Rusponi, F. Buatier de Mongeot, C.
Boragno, and U. Valbusa, J. Phys.: Condens. Matter 13 (2001) 5875.

\item[{\lbrack 12]}]  J. H. Jeffries, J. -K. Zuo, and M. M. Craig, Phys.
Rev. Lett. 76 (1996) 4931.

\item[{\lbrack 13]}]  S. Wei, B. Li, T. Fujimoto, and I. Kojima, Phys. Rev.
B 58 (1998) 3605.

\item[{\lbrack 14]}]  S. M. Rossnagel and R. S. Robinson, J. Vac. Sci.
Technol. 20 (1982) 195.

\item[{\lbrack 15]}]  G. Carter, J. Phys. D: Appl. Phys.34 (2001) R1-R22.

\item[{\lbrack 16]}]  G. K. Wehner, Appl. Phys. Lett. 43 (1983) 366.

\item[{\lbrack 17]}]  G. Carter and V. Vishnyakov, Phys. Rev. B 54
(1996)17647.

\item[{\lbrack 18]}]  Z. F. Ziegler, IBM Research, SRIM-2000.40 (PC
version), Yorktown Heights, NY, 1999.

\item[{\lbrack 19]}]  P. Karmakar and D. Ghose, to be published.

\item[{\lbrack 20]}]  S. Habenicht, K. P. Lieb, J. Koch and A. D. Wieck,
Phys. Rev. B 65 (2002)115327.

\item[{\lbrack 21]}]  G. Carter, V. Vishnyakov and M. J. Nobes, Nucl.
Instrum. and Meth. B 115 (1996) 440.

\pagebreak
\end{itemize}

\section*{Figure captions}

\smallskip

Fig. 1. AFM images of the unbombarded and $16.7$ keV $Ar^{+}$ sputtered $Co$%
, $Cu$, $Ag$, $Pt$ and $Au$ surfaces at different angles of incidence $%
\theta $ as indicated. The bombarding dose $\phi $ for $Co$, $Cu$, $Ag$, and 
$Au$ is $1\times 10^{17}$ $ions/cm^{2}$, while that for $Pt$ is $5\times
10^{16}$ $ions/cm^{2}$. The ion beam direction is from the bottom to the
top.\smallskip

Fig. 2. (a) AFM image of the rippled surface on $Au$ film after $16.7$ keV $%
Ar^{+}$ ion sputtering at $\theta =80^{0}$ and $\phi =1\times 10^{17}$ $%
ions/cm^{2}$; the ion beam direction is from the bottom to the top. (b)
showing the corresponding 2D-autocorrelation function. (c) showing the
1D-autocorrelation function along the marked line in (b) in order to
determine $\Lambda $.

\smallskip Fig. 3. The ripple wavelength $\Lambda $ versus the substrate
element after $16.7$ keV $Ar^{+}$ ion sputtering at $\theta =80^{0}$ and $%
\phi =1\times 10^{17}$ $ions/cm^{2}$.

Fig. 4. AFM image of the ripples formed on a $Si(100)$ surface after $16.7$
keV $Ar^{+}$ ion sputtering at $\theta =60^{0}$ and $\phi =5\times 10^{17}$ $%
ions/cm^{2}$. The ion beam direction is from the bottom to the top.

\end{document}